%Paper: cond-mat/9406056
%From: Mark Srednicki <mark@tpau.physics.ucsb.edu>
%Date: Mon, 13 Jun 94 18:24:28 -0700

%%%%%%% CUT HERE; all macros included; no figures %%%%%%%%%%%%%%%%%%%%%%%%%%%
\magnification1200
\overfullrule=0pt
\parskip=4pt plus 2pt
\newcount\firstpageno
\firstpageno=2
\footline={\ifnum\pageno<\firstpageno{\hfil}
           \else{\hfil\folio\hfil}\fi}
\def\references{\frenchspacing \parindent=0pt \leftskip=0.8truecm
   \rightskip=0truecm \parskip=4pt plus 2pt \everypar{\hangindent=\parindent}}
\def\endreferences{\vskip0pt}
\def\refstyleprl{
 \gdef\r##1{$^{##1}$}	         			% Reference in text []
 \gdef\refis##1{\indent\hbox to 0pt{\hss{##1}.~~}}     	% Ref list numbers
 \gdef\citerange##1##2##3{$^{\cite{##1}-\setbox0=\hbox{\cite{##2}}
                             \cite{##3}}$}}
\def\frac#1#2{{\textstyle{#1 \over #2}}} 
  \def\ss{\scriptstyle}
\def\sss{\scriptscriptstyle}
\def\gtwid{\mathrel{\raise.3ex\hbox{$>$\kern-.75em\lower1ex\hbox{$\sim$}}}}
\def\ltwid{\mathrel{\raise.3ex\hbox{$<$\kern-.75em\lower1ex\hbox{$\sim$}}}}
\refstyleprl
\catcode`@=11
\newcount\r@fcount \r@fcount=0
\newcount\r@fcurr
\immediate\newwrite\reffile
\newif\ifr@ffile\r@ffilefalse
\def\w@rnwrite#1{\ifr@ffile\immediate\write\reffile{#1}\fi\message{#1}}
\def\writer@f#1>>{}
\def\referencefile{%			  Stuff to write .REF file
  \r@ffiletrue\immediate\openout\reffile=\jobname.ref%
  \def\writer@f##1>>{\ifr@ffile\immediate\write\reffile%
    {\noexpand\refis{##1} = \csname r@fnum##1\endcsname = %
     \expandafter\expandafter\expandafter\strip@t\expandafter%
     \meaning\csname r@ftext\csname r@fnum##1\endcsname\endcsname}\fi}%
  \def\strip@t##1>>{}}

\def\citeall#1{\xdef#1##1{#1{\noexpand\cite{##1}}}}
\def\cite#1{\each@rg\citer@nge{#1}}	% Variable # of args, separated by ","
\def\each@rg#1#2{{\let\thecsname=#1\expandafter\first@rg#2,\end,}}
\def\first@rg#1,{\thecsname{#1}\apply@rg}	% each@ag is a general purpose
\def\apply@rg#1,{\ifx\end#1\let\next=\relax%	  variable no. of arg. macro.
\else,\thecsname{#1}\let\next=\apply@rg\fi\next}% args separated by commas
\def\citer@nge#1{\citedor@nge#1-\end-}	% Check for M-N range (M and N numbers)
\def\citer@ngeat#1\end-{#1}
\def\citedor@nge#1-#2-{\ifx\end#2\r@featspace#1 % Single argument
  \else\citel@@p{#1}{#2}\citer@ngeat\fi}	% M-N range of arguments
\def\citel@@p#1#2{\ifnum#1>#2{\errmessage{Reference range #1-#2\space is bad.}%
    \errhelp{If you cite a series of references by the notation M-N, then M and
    N must be integers, and N must be greater than or equal to M.}}\else%
 {\count0=#1\count1=#2\advance\count1 by1\relax\expandafter\r@fcite\the\count0,
  \loop\advance\count0 by1\relax%	  Loop from M to N
    \ifnum\count0<\count1,\expandafter\r@fcite\the\count0,%
  \repeat}\fi}
\def\r@featspace#1#2 {\r@fcite#1#2,}	% Eat spaces at beginning or end of arg
\def\r@fcite#1,{\ifuncit@d{#1}%		  Cite individual reference
    \newr@f{#1}%
    \expandafter\gdef\csname r@ftext\number\r@fcount\endcsname%
                     {\message{Reference #1 to be supplied.}%
                      \writer@f#1>>#1 to be supplied.\par}%
 \fi%
 \csname r@fnum#1\endcsname}
\def\ifuncit@d#1{\expandafter\ifx\csname r@fnum#1\endcsname\relax}%
\def\newr@f#1{\global\advance\r@fcount by1%
    \expandafter\xdef\csname r@fnum#1\endcsname{\number\r@fcount}}
\let\r@fis=\refis			% Save old \refis, redefine
\def\refis#1#2#3\par{\ifuncit@d{#1}%      Use two params #2 #3 to strip blank
   \newr@f{#1}%
   \w@rnwrite{Reference #1=\number\r@fcount\space is not cited up to now.}\fi%
  \expandafter\gdef\csname r@ftext\csname r@fnum#1\endcsname\endcsname%
  {\writer@f#1>>#2#3\par}}
\def\ignoreuncited{%   redefine \refis if ignoring uncited references
   \def\refis##1##2##3\par{\ifuncit@d{##1}%
     \else\expandafter\gdef\csname r@ftext\csname r@fnum##1\endcsname\endcsname
     {\writer@f##1>>##2##3\par}\fi}}
\def\r@ferr{\endreferences\errmessage{I was expecting to see
\noexpand\endreferences before now;  I have inserted it here.}}
\let\r@ferences=\references
\def\references{\r@ferences\def\endmode{\r@ferr\par\endgroup}}
\let\endr@ferences=\endreferences
\def\endreferences{\r@fcurr=0%		  Save old \endreferences, redefine
  {\loop\ifnum\r@fcurr<\r@fcount%	  Loop over refnum and produce text
    \advance\r@fcurr by 1\relax\expandafter\r@fis\expandafter{\number\r@fcurr}%
    \csname r@ftext\number\r@fcurr\endcsname%
  \repeat}\gdef\r@ferr{}\endr@ferences}
\let\r@fend=\endpaper\gdef\endpaper{\ifr@ffile
\immediate\write16{Cross References written on []\jobname.REF.}\fi\r@fend}
\catcode`@=12
\citeall\r		%
\def\cd{\!\cdot\!} \def\a{\alpha} \def\b{\beta} \def\d{\delta}
\def\p{{\bf p}} \def\P{{\bf P}} 
\def\X{{\bf X}} \def\tpsi{{\widetilde\psi}} \def\paa{\Phi_{\a\a}}
\def\pab{\Phi_{\a\b}} \def\fmb{f_{\rm\sss MB}}
\def\intp#1{\int d^3p_{#1} \ldots d^3p_N\ }
\def\apriori{{\it a priori\/\ }}
\def\psphere{P-sphere}
\baselineskip=14pt
\rightline{cond-mat/9406056}
\rightline{UCSB--TH--94--17}
\rightline{June 1994}
\font \bigbf=cmbx12 scaled \magstep2
\vskip 3pt plus 0.4fill
\centerline{\bigbf Quantum Chaos and Statistical Mechanics}
\vskip 3pt plus 0.2fill
\font \smcap=cmcsc10
\centerline{\smcap Mark Srednicki}
\vskip 3pt plus 0.1fill
\centerline{\sl Department of Physics\/}
\centerline{\sl University of California\/}
\centerline{\sl Santa Barbara, CA 93106\/}
\centerline{\tt mark@tpau.physics.ucsb.edu}
\vskip 3pt plus 0.2fill
ABSTRACT:~~We briefly review the well known connection between classical chaos
and classical statistical mechanics, and the recently discovered connection
between quantum chaos and quantum statistical mechanics.
\vskip 3pt plus 0.2fill
\centerline{\sl Talk given at the Conference on}
\centerline{\sl Fundamental Problems in Quantum Theory}
\centerline{\sl Baltimore, June 18--22, 1994}
\vskip 3pt plus 0.4fill
\eject
%\raggedright

Consider a dilute gas of hard spheres in a box with hard walls.
Give the spheres some arbitrary initial distribution of momenta
(and positions).  Classically, after a few mean free times have passed,
we expect that the distribution of momenta will be given by the
Maxwell--Boltzmann (MB) formula,
$$\fmb(\p) = (2\pi mkT)^{-3/2}\exp(-\p^2\!/2mkT)\;,              \eqno(1)$$
where the temperature $T$ is given in terms of the conserved total energy $U$
by the \break
ideal-gas relation $U=\frac32 NkT$.

To see why this should be so, first note that the hamiltonian is simply
$$H = {1\over 2m}\sum_{i=1}^N \p_i^2 = {1\over 2m}\P^2\;,       \eqno(2)$$
where $\P$ is a vector with $3N$ components.  Since $H$ takes on the constant
value $U$, the allowed values of $\P$ form a sphere which we will call
the \psphere.  Suppose we now choose $\P$ ``at random.''  For this to be
a meaningful statement, we must have a measure which tells us which sets
of $\P$'s are equally likely {\it a priori}.  The obvious choice is to assign
equal \apriori probabilities to equal areas on the \psphere.  Then if we choose
$\P$ at random with respect to this measure, the probability that our choice
makes an angle between $\theta$ and $\theta+d\theta$ with respect to any
particular axis is simply
$$\eqalignno{
f(\theta)\,d\theta  &\sim (\sin\theta)^{3N-2}\,d\theta \cr
\noalign{\medskip}
                    &\sim (\sin\theta)^{3N-3}\,d\cos\theta \cr
\noalign{\medskip}
                    &\sim (1-\cos^2\theta)^{(3N-3)/2}\,d\cos\theta \;. & (3)
                                                                      \cr} $$
If we now identify $(2mU)^{1/2}\cos\theta$ as, say, the value of $p_{1z}$
(the $z$ component of the first particle's momentum), we find
$$\eqalignno{
f(p_{1z})\,dp_{1z} &\sim (1-p_{1z}^2/2mU)^{(3N-3)/2}\,dp_{1z} \cr
\noalign{\medskip}
                   &\sim \exp(-p_{1z}^2/2mkT)\,dp_{1z}  \;,&(4) \cr}$$
where in the second line we have set $U=\frac32 NkT$ and taken the large-$N$
limit.  Thus we have recovered the MB distribution for $p_{1z}$.  Now consider
the probability distribution for $p_{1y}$ when $p_{1z}$ is fixed; it is
given by the first line of (4) with $3N$ replaced by $3N-1$
(since there is one less coordinate when $p_{1z}$ is fixed)
and $2mU$ replaced by $2mU-p_{1z}^2$.  In the large-$N$ limit,
we can neglect $p_{1z}^2$ compared to $2mU$, and we find the MB
distribution for $p_{1y}$.  In similar fashion, we get the MB distribution
for any $n$ components of $\P$ as long as $n\ll N$.

Now our task is to justify the assumption that equal areas on the
\psphere\ are equally likely {\it a priori}.  That such a justification
is needed can be seen by considering how we would go about filling a
real box with a real gas (say, helium).  If the box did not already have some
sort of valve on it, we would install one, and pump the air out through it.
Then we would close the valve, attach it to a tank of helium with a hose,
and open the valve.  The helium atoms would rush in, moving preferentially
in the direction parallel to the hose.  Thus their initial distribution of
momenta would be strongly anisotropic.   This is in sharp contrast to the
prediction of the equal-area measure, which tells us that we will
find a thermal, isotropic distribution.  Clearly, then, the equal-area
measure has nothing to do with how we put real gases in real boxes,
and so we must seek its justification elsewhere.

That justification comes from Sinai's theorem\r{sinai},
which states that a box of hard spheres is a {\it chaotic\/} system.
The meaning of this statement in the present context is simple.
Start off with arbitrary initial momenta and positions; the momenta
can be as nonthermal as you like.  (Actually, we must exclude a set
of measure zero; for example, it is possible to set up initial conditions
such that no two hard spheres ever collide, in which case the following
discussion obviously does not apply.)
Wait a few mean free times, and then note the current location of $\P$
on the \psphere.  Continue this procedure, keeping track of the location
of $\P$ each time.  Chaos implies that this sequence of $\P$'s appears to be
chosen at random with respect to the equal-area measure.

We are done.  Even if we started off with a $\P$ representing a strongly
anisotropic distribution, the next $\P$ will appear to be chosen ``at random,''
and so predicts a thermal distribution for the individual momenta.

So much for classical mechanics.  What about quantum mechanics?

Now we have a completely different problem\r{sred}.
The $N$-particle Schr\"odinger equation can always be solved by going to the
energy eigenstate basis: $H|\alpha\rangle=U_\alpha|\alpha\rangle$.
The hamiltonian is given by~(2), supplemented by the boundary condition
that the energy eigenfunctions $\psi_\alpha(\X)$ vanish whenever one of the
hard spheres touches a wall of the box, or whenever two hard spheres touch
each other.  The wave function in momentum space at time $t$ is then
$$\tpsi(\P,t) = \sum_\a C_\a\exp(-iU_\a t/\hbar)\,\tpsi_\a(\P)\;,  \eqno(5)$$
where the $C_\a$'s specify the initial state.  The probability that the
first particle has momentum $\p_1$ at time $t$ is found by squaring the
wave function and integrating over all momenta but the first:
$$\eqalignno{
f(\p_1,t) &= \intp2 \bigl|\tpsi(\P,t)\bigr|{}^2 \cr
\noalign{\medskip}
          &= \sum_{\a\b}C_\a^* C_\b\,e^{i(U_\a-U_\b)t/\hbar}
             \intp2 \tpsi^*_\a(\P)\tpsi_\b(\P)\cr
\noalign{\medskip}
          &= \sum_{\a\b}C_\a^* C_\b\,e^{i(U_\a-U_\b)t/\hbar}
                \;\pab(\p_1)\;.                                   &(6)\cr}$$
In the last line we have introduced
$$\Phi_{\a\b}(\p_1)\equiv\intp2 \tpsi_\a^*(\P)\tpsi_\b(\P)\;,\eqno(7)$$
which obeys the normalization condition
$$\int d^3p_1\,\Phi_{\a\b}(\p_1)=\d_{\a\b}\;.               \eqno(8)$$
If we symmetrize or antisymmetrize each $\tpsi_\a(\P)$ on exchange of any
two $\p_i$'s to reflect Bose--Einstein (BE) or Fermi--Dirac (FD) statistics,
then $f(\p_1,t)$ is independent of which $\p_i$ we choose.

On physical grounds, we expect that $f(\p_1,t)$ should be the MB (or BE or FD)
distribution for any time $t$ greater than a few mean free times.  It is not
obvious how this can occur.  Consider, for example, what happens if we take
the infinite time average of $f(\p_1,t)$:
$$\lim_{\tau\to\infty}{1\over\tau}\int_0^\tau dt\;f(\p_1,t)
  = \sum_\a\bigl|C_\a\bigr|^2 \,\paa(\p_1)\;.                 \eqno(9)$$
The infinite time average is obviously not something we can actually observe,
but theoretically, if anything is going to be thermal, this is it.  The
problem is that the $C_\a$'s are essentially arbitrary, so how can we
possibly get the MB distribution?

There is only one way:  each $\paa(\p_1)$ must {\it individually\/} be equal
to the MB (or BE or FD) distribution at a temperature $T_\a$ which is given
(at least approximately) by the ideal-gas relation $U_\a = \frac32 NkT_\a$.
We call this hypothesis {\it eigenstate thermalization\/}.  If eigenstate
thermalization is valid, then (9) will indeed be a thermal distribution as
long as the uncertainty in the total energy is much less than its expectation
value.

Furthermore, if $\pab(\p_1)$ is always sufficiently small whenever $\a\ne\b$,
then the $\a\ne\b$ terms in (6) will usually make a negligible contribution,
and $f(\p_1,t)$ will be a thermal distribution at most times $t$, without any
time averaging at all.  However, if the magnitudes and phases of the $C_\a$'s
are carefully chosen, then we can ``line up'' the $\pab(\p_1)$'s so as to get
any $f(\p_1,t)$ that we might want at any one particular time (say, $t=0$).
Afterward, however, as we see in (6), the phases will change in the usual
manner; the carefully contrived coherence among the various $\pab(\p_1)$'s
will be destroyed, and we will again find a thermal distribution for $\p_1$.

I find this to be a clear and satisfying explanation for the validity of
quantum statistical mechanics, at least in this particular problem, even
without any further evidence in favor of it.  However, there is more to be
said:  a very strong case can be made for the two necessary ingredients---the
thermal nature of $\paa(\p_1)$ and the smallness of $\pab(\p_1)$---based
on the theory of quantum chaos.

Quantum chaos is the study of quantum systems whose classical counterparts
are chaotic.  The result we will need is known as Berry's
conjecture\citerange{berry77b}{berry81}{berry91}.
As its name implies, Berry's conjecture is as yet unproved, but there is
significant numerical evidence (reviewed in [2]) in support of it.

Berry's conjecture has
two parts.  Part one says that the energy eigenfunctions of a bounded,
isolated quantum system which is classically chaotic appear to be
gaussian random variables, in the sense that
$$\lim_{\a\to\infty} \int d\X\ \psi_\a(\X+\X_1)\ldots\psi_\a(\X+\X_n)
  = \sum_{\rm pairs} J(\X_{i_{\ss 1}}-\X_{i_{\ss 2}})
                 \ldots J(\X_{i_{\ss n-1}}-\X_{i_{\ss n}})   \;. \eqno(10)$$
Here the integration measure is normalized so that $\int d\X=1$,
and the sum is over all possible ways to pair up the $\X_i$'s;
if $n$ is odd the result is zero.
Part two says that the correlation function $J(\X)$ is given by
$$J(\X)\sim\int d\P\ \exp(i\P\cd\X/\hbar)\ \delta\bigl(H(\P,\X)-U_\a\bigr)\;,
                                                                 \eqno(11)$$
where $\delta(x)$ is the Dirac delta function, and $J({\bf 0})=1$.

It is straightforward to show that Berry's conjecture gives us
the two necessary ingredients for quantum statistical mechanics.
First, we find that when $\a\ne\b$,
$\pab(\p_1)$ is exponentially small in the number of particles $N$.
More importantly, we find eigenstate thermalization:  $\paa(\p_1)$ is given by
the MB or BE or FD distribution (depending on whether we use nonsymmetric,
completely symmetric, or completely antisymmetric energy eigenfunctions), plus
corrections which depend on the specific energy eigenfunction but which are
exponentially small in $N$.  To derive this, the gas must be dilute; there are
also other, hard-to-compute corrections to $\paa(\p_1)$ due to the finite
radii of the hard spheres.  We expect these to reproduce the usual hard-sphere
corrections to ideal-gas behavior, but this remains to be demonstrated.
Another important unsettled issue is how high the energy needs to be before
(10) is sufficiently accurate.  A naive estimate is $\lambda_\a\ltwid a$,
where $\lambda_\a=(2\pi\hbar^2/mkT_\a)^{1/2}$ is a typical particle's
de Broglie wavelength, and $a$ is the hard-sphere radius.
With $a$ in angstroms and $m$ in amu,
this condition becomes $T_\a\gtwid(300/ma^2)\;$Kelvin.

To summarize, the appearance of a thermal distribution of momenta in
an isolated, bounded quantum system of many particles can only be understood
if each energy eigenfunction individually predicts a thermal probability for
the momentum of each constituent particle, and if overlaps of different energy
eigenfunctions are sufficiently small when one particle's momentum is left
unintegrated.  Both these statements can be derived as consequences of
Berry's conjecture, which is expected to hold only for quantum systems
whose classical counterparts are chaotic.  Thus the well known
connection between classical statistical mechanics and classical chaos
is now seen to be mirrored by an analogous connection between
quantum statistical mechanics and quantum chaos.

\vskip0.2in

I would like to thank Edouard Brezin, John Cardy, Marty Halpern,
and Walter Kohn for helpful discussions.
This work was supported in part by NSF Grant PHY--91--16964.

\vskip0.3in
\noindent\centerline{References}
\vskip0.2in

\references

\refis{sinai}Ya. G. Sinai, Ups. Mat. Nauk 25, 137 (1970)
[Russ. Math. Surv. 25, 137 (1970)].

\refis{sred}M. Srednicki, cond-mat/9403051, Phys. Rev. E, in press.

\refis{berry77b}M. V. Berry, J. Phys. A 10, 2083 (1977).

\refis{berry81}M. V. Berry, in {\sl Les Houches XXXVI, Chaotic Behavior of
Deterministic Systems}, G.~Iooss, R. H. G. Helleman, and R. Stora, eds.
(North-Holland, Amsterdam, 1983).

\refis{berry91}M. V. Berry, in {\sl Les Houches LII,
Chaos and Quantum Physics}, M.-J. Giannoni, A.~Voros, and J. Zinn-Justin, eds.
(North--Holland, Amsterdam, 1991).

\endreferences
\end